\documentclass[prl,letterpaper,twocolumn]{revtex4-1}
\usepackage[utf8x]{inputenc}
\usepackage{amsmath}
\usepackage{amssymb}
\usepackage{graphicx}
\usepackage{dcolumn}
\usepackage{bm}
\usepackage{subfigure}

\usepackage{color}

\newcommand{\abs}[1]{\left| #1 \right|}

\newcommand{\steveB}[1]{\mathbf{#1}}
\newcommand{\steveBk}{\steveB{k}}

\begin{document}
\title{\bf Filling-Enforced Magnetic Dirac Semimetals in Two Dimensions}
\author{Steve M. Young}
\affiliation{Center for Computational Materials Science, Naval Research Laboratory, Washington, D.C. 20375, USA}
\author{Benjamin J. Wieder}
\affiliation{Nordita, Center for Quantum Materials, KTH Royal Institute of Technology and Stockholm University, Roslagstullsbacken 23, SE-106 91 Stockholm, Sweden}
\begin{abstract}
Filling-enforced Dirac semimetals, or those required at specific fillings by the combination of crystalline and time-reversal symmetries, have been proposed in numerous materials.  However, Dirac points in these materials are not generally robust against breaking or modifying time-reversal symmetry.  We present a new class of two-dimensional Dirac semimetal protected by the combination of crystal symmetries and a special, antiferromagnetic time-reversal symmetry.  Systems in this class of magnetic layer groups, while having broken time-reversal symmetry, still respect the operation of time-reversal followed by a half-lattice translation. In contrast to 2D time-reversal-symmetric Dirac semimetal phases, this magnetic Dirac phase is capable of hosting just a single isolated Dirac point at the Fermi level, one that can be stabilized solely by symmorphic crystal symmetries.  We find that this Dirac point represents a new quantum critical point, existing at the boundary between Chern insulating, antiferromagnetic topological crystalline insulating, and trivial insulating phases, and we discuss its relationship with condensed matter fermion doubling theorems.  We present density functional theoretic calculations which demonstrate the presence of these 2D magnetic Dirac points in FeSe monolayers and discuss the implications for engineering quantum phase transitions in these materials.  
\end{abstract}
\maketitle

Since the discovery of graphene~\cite{Castro_Neto_2009,GrapheneDirac1,semenoff,meleDirac} and the recognition of the unique role of its Dirac cones in transport~\cite{Novoselov_2005,kleinTunneling1,kleinTunneling2} and quantum criticality~\cite{Kane05p226801,Kane05p146802}, there has been an ongoing effort to reproduce aspects of Dirac semimetal physics in new materials and to predict new variants.  Through this search, many new semimetallic phases have been predicted, characterized, and, in some cases, observed in real materials, including phases hosting 3D Dirac, Weyl, Double Dirac, Spin-1 Weyl, or line nodes near the Fermi energy~\cite{Burkov_2011, Wan_2011, Halasz_2012,Liu_2014_3, Weng_2015,Xu_2015,Wang_2012,Wang_2013,Borisenko_2014,Liu_2014_1,Liu_2014_2,Lv_2015,SteveDirac,Soluyanov_2015,Soluyanov_2016,Pixley2016,Venderbos2016,DDP,NewFermions}. 

In these semimetallic phases, the nodal features are stabilized by the combination of time-reversal and spatial symmetries.  In particular, for the phases protected by nonsymmorphic symmetries, or those invariant under the combination of a point group operation and a fractional lattice translation, certain nodal features are always present at space-group-specific fillings~\cite{ashvin,WPVZ,WiederLayers}.  These semimetals, known as ``filling-enforced semimetals,'' are prevented from being insulators at these fillings by the combination of Kramers theorem and nonsymmorphic symmetries and, therefore, display bands inseparably bound together in space-group-specific numbers.  For example, for the simple 2D four-band models previously presented in Ref.~\citenum{Steve2D}, glides and twofold screws forced bands to tangle together in groups of four, such that at filling $\nu=2$ the system always displayed Weyl or Dirac points.  Unlike the Dirac points in band-inversion semimetals Na$_3$Bi and Cd$_2$As$_3$~\cite{Wang_2012,Wang_2013}, filling-enforced nodal features can be found in all time-reversal-symmetric materials~\cite{KramersWeyl}.

In this Letter, we present the first examples of filling-enforced Dirac semimetals in systems with magnetic symmetries.  We find that the combination of three-dimensional layer group crystal symmetries and an antiferromagnetic time-reversal symmetry protects a single bulk Dirac point in a two-dimensional crystal, and we present four-band tight-binding models demonstrating this physics.  Unlike in time-reversal-symmetric Dirac semimetals, a single magnetic Dirac point is permitted to exist as the only feature at the Fermi energy.  Furthermore, unlike the antiferromagnetic Dirac points in Ref~\citenum{ShouchengAF}, which are topological objects created through band-inversion transitions, this 2D magnetic Dirac point is filling enforced: it cannot be gapped without lowering the symmetry of the particular magnetic layer group that protects it.

Further, we show that this magnetic Dirac point, like its time-reversal-symmetric relative, represents the quantum critical point between topologically distinct insulating phases. For bulk perturbations that preserve the antiferromagnetic time-reversal operation, this magnetic Dirac semimetal sits at the quantum phase boundary between a trivial insulator, a Chern-trivial antiferromagnetic topological crystalline insulator, and a nontrivial Chern insulator with winding $C=\pm 1$.

We present density functional theoretic (DFT) calculations demonstrating the presence of these magnetic Dirac points in FeSe monolayers, and discuss the implications for engineering topological phase transitions in magnetic Dirac semimetals.  Finally, we discuss the stability of this new magnetic Dirac fermion in the context of disorder, interactions, and condensed matter fermion doubling theorems.

In time-reversal-symmetric, filling-enforced Dirac semimetals, Dirac nodes at time-reversal-invariant momenta (TRIMs) are protected by the algebraic relationship between two spatial symmetries and time-reversal symmetry~\cite{Steve2D,WiederLayers}.  Specifically, at the $\steveB{k}\cdot\steveB{p}$ level, a fourfold point degeneracy may be protected by two spatial operations $A$ and $B$ and an antiunitary operation $\bar{\mathcal{T}}$ whose irreducible representations satisfy the algebra

\begin{eqnarray}
\left\{A,B\right\}&=&\left[A,\bar{\mathcal{T}}\right]=\left[B,\bar{\mathcal{T}}\right]=0, \nonumber \\
\quad A^2&=&\pm B^2=-\bar{\mathcal{T}}^2=+ 1.
\label{eq:alg}
\end{eqnarray}

As magnetically ordered systems can still possess a time-reversal-like antiunitary symmetry, they are also capable of satisfying these relations, and, in fact, may do so utilizing an expanded set of crystal symmetry operations.  Specifically, when $\bar{\mathcal{T}}$ is plain time-reversal symmetry $\mathcal{T}$, the above relations may only be satisfied if at least one of $A$ or $B$ is a twofold nonsymmorphic operation~\cite{WiederLayers}.  However, at $\bar{\mathcal{T}}$-invariant momenta in magnetically ordered systems, or those for which $\bar{\mathcal{T}}=\left\{\mathcal{T} | \steveB{t} \right\}$, where $\steveB{t}$ is a fractional translation, commutation relations between $\bar{\mathcal{T}}$ and spatial symmetries may be altered, allowing Eq.~\eqref{eq:alg} to be satisfied using only symmorphic symmetries. 

\begin{figure}
\centering
\subfigure[]{\includegraphics[bb=-30 -30 220 220,height=3.5cm]{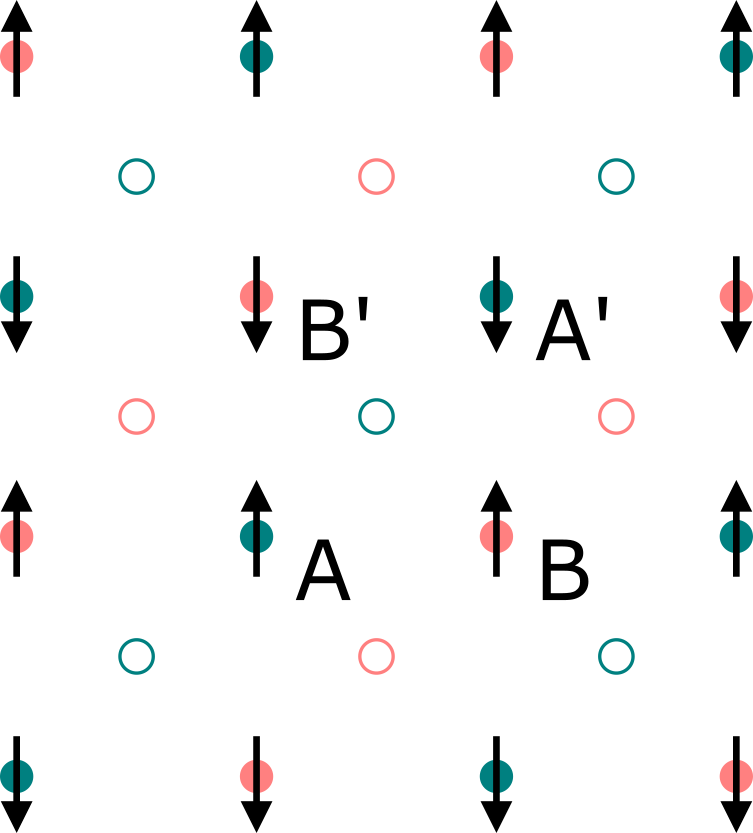}\label{fig:lattice_ic2z}}
\subfigure[]{\includegraphics[bb=50 0 600 400,height=3.5cm]{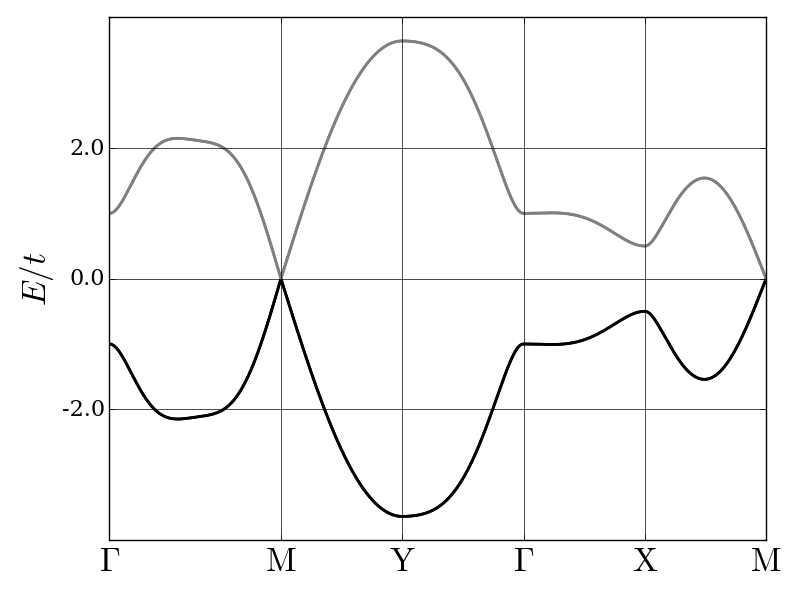}\label{fig:bands_ic2z}}

\caption{\subref{fig:lattice_ic2z} The lattice with $\mathcal{I}$, $\{M_{z} | 0 \frac{1}{2} \}$, and $\bar{\mathcal{T}}=\{\mathcal{T}| \frac{1}{2}  \frac{1}{2}\}$ with spin alignment $\pm\hat{y}$. Red and green dots indicate sites above and below the plane.  \subref{fig:bands_ic2z} The band structure generated by the tight-binding model with these symmetries [Eq.~\eqref{eq:tb1}. Bands are twofold-degenerate by the combination of $\mathcal{I}$ and $\bar{\mathcal{T}}$.  Pictured are the top four bands of an eight-band model, which are split from the bottom bands by a very large antiferromagnetic interaction.  The symmetries of this magnetic layer group necessitate that groups of four bands meet in Dirac points at $M$ for fillings $\nu\in 4\mathbb{Z} + 2$.}
\label{fig:ic2z}
\end{figure}

To have a composite time-reversal-like symmetry $\bar{\mathcal{T}}$ with a  fractional lattice translation, a system must be composed of sites that, while internally time-reversal-broken, have time-reversed partners elsewhere in the unit cell.  The simplest example of this is a two-site antiferromagnet, where the up spins on the $A$ sites are the time-reverses of the down spins on the $B$ sites.  To construct a model with this symmetry, we first consider systems with four sublattices of $s$-orbitals, for a total of eight bands.  Then, we turn on an antiferromagnetic potential, assumed to be much stronger than other hopping and energy terms, such that the system splits into two effectively four-band systems, each with one spin per sublattice.  The subsystem above the plane can, therefore, be described using two pairs of sublattices $A,B$.   Each pair individually respects $\bar{\mathcal{T}}$, and the two pairs are related to one another by additional spatial symmetries, as shown in Fig.~\ref{fig:lattice_ic2z}, for which $\steveB{t}= \left(  \frac{1}{2}  \frac{1}{2} \right)$.  In this model, there are also additional spatial symmetries inversion $\mathcal{I}$ and glide reflection $M_{z}$.  Representing the $A$ or $B$ degrees of freedom by $\sigma$ and the prime or nonprime degrees of freedom by $\tau$, the $\steveBk\cdot\steveB{p}$ model of the $M$ point ($k_{x}=k_{y}=\pi$) reads

\begin{flalign*}
&\mathcal{I}=i\tau_y,M_z=i\tau_x\sigma_y,\bar{\mathcal{T}}=i\tau_z\sigma_yK\\
H_{\rm M}= &\left[t_0\tau_x+\left(t_2^{\rm SO}+t_3^{\rm SO}\right)\tau_z\sigma_z+\left(t_3^{\rm SO}+t_4^{\rm SO}\right)\tau_z\sigma_x\right] k_x\\
&- \left[\left(t_2^{\rm SO}-t_3^{\rm SO}\right)\tau_z\sigma_z+\left(t_3^{\rm SO}-t_4^{\rm SO}\right)\tau_z\sigma_x\right]k_y
\end{flalign*}

and can be generated by the tight-binding model

\begin{flalign}
\mathcal{H}
=&t_0\cos\left(\frac{k_x}{2}\right)\tau_x\nonumber\\
&+\left[t_1^{\rm SO}\sin\left(k_x-k_y\right)+t_2^{\rm SO}\sin\left(k_x+k_y\right)\right]\tau_z\sigma_z\nonumber\\
&+\left[t_3^{\rm SO}\sin\left(\frac{k_x-k_y}{2}\right)+t_4^{\rm SO}\sin\left(\frac{k_x+k_y}{2}\right)\right]\tau_z\sigma_x\label{eq:tb1}
\end{flalign}
for which we have enumerated all symmetry-allowed terms up to second-nearest-neighbor hopping.  

If these two spatial symmetries were combined with regular time-reversal-symmetry $\mathcal{T}$, Dirac points at $M$ and $Y$ would result, as shown in Ref.~\onlinecite{Steve2D}.  However, for an antiferromagnet with $\bar{\mathcal{T}}$ symmetry there is only a Dirac point at $M$, as in the little group at $Y$, the translation $\steveB{t}$ anticommutes with both spatial operations and the algebra in Eq.~\eqref{eq:alg} is no longer satisfied.  

\begin{figure}
\centering
\subfigure[]{\includegraphics[bb=-39 0 363 324,width=4.0cm]{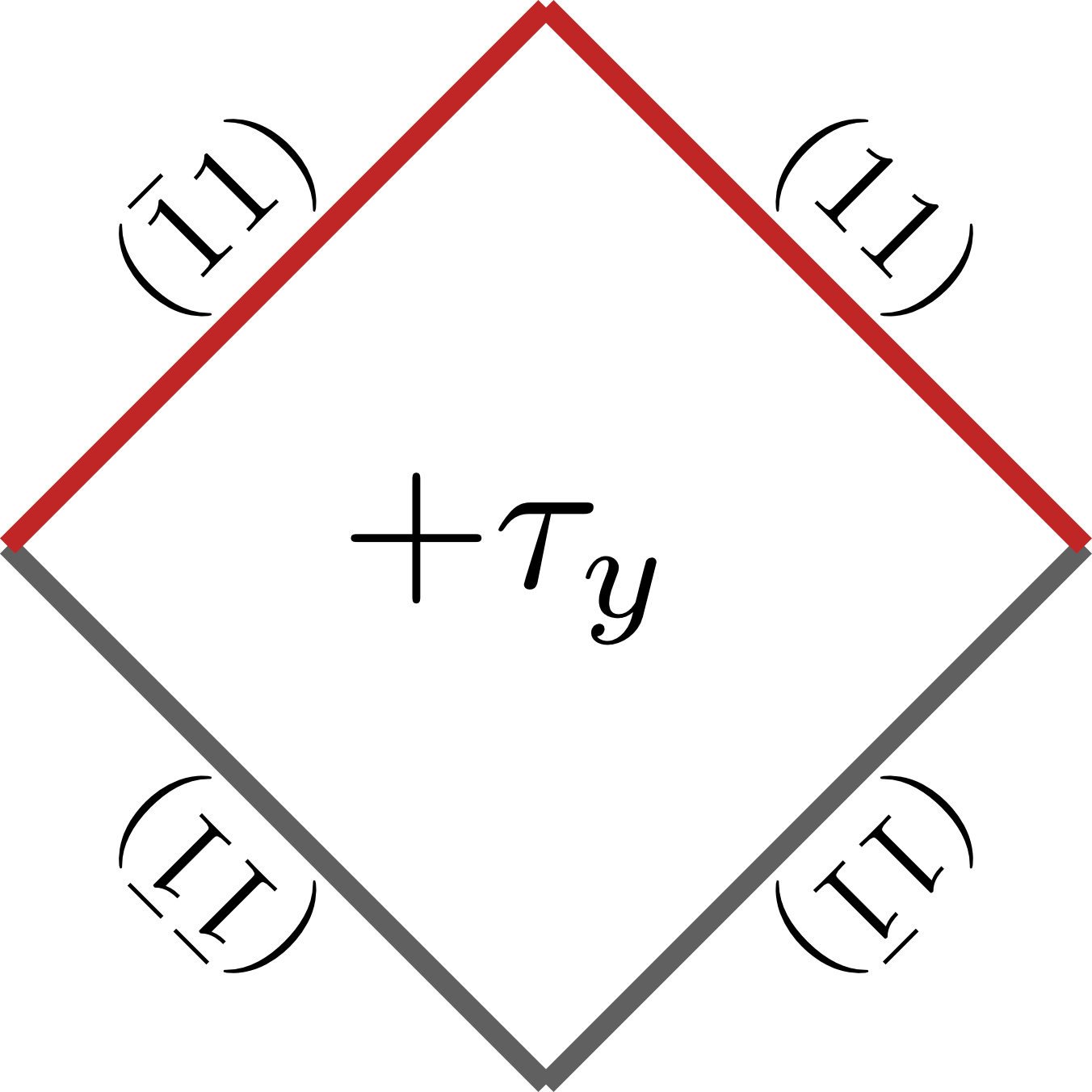}\label{fig:ic2z_surf_a1}}
\subfigure[]{\includegraphics[bb=-39 0 363 324,width=4.0cm]{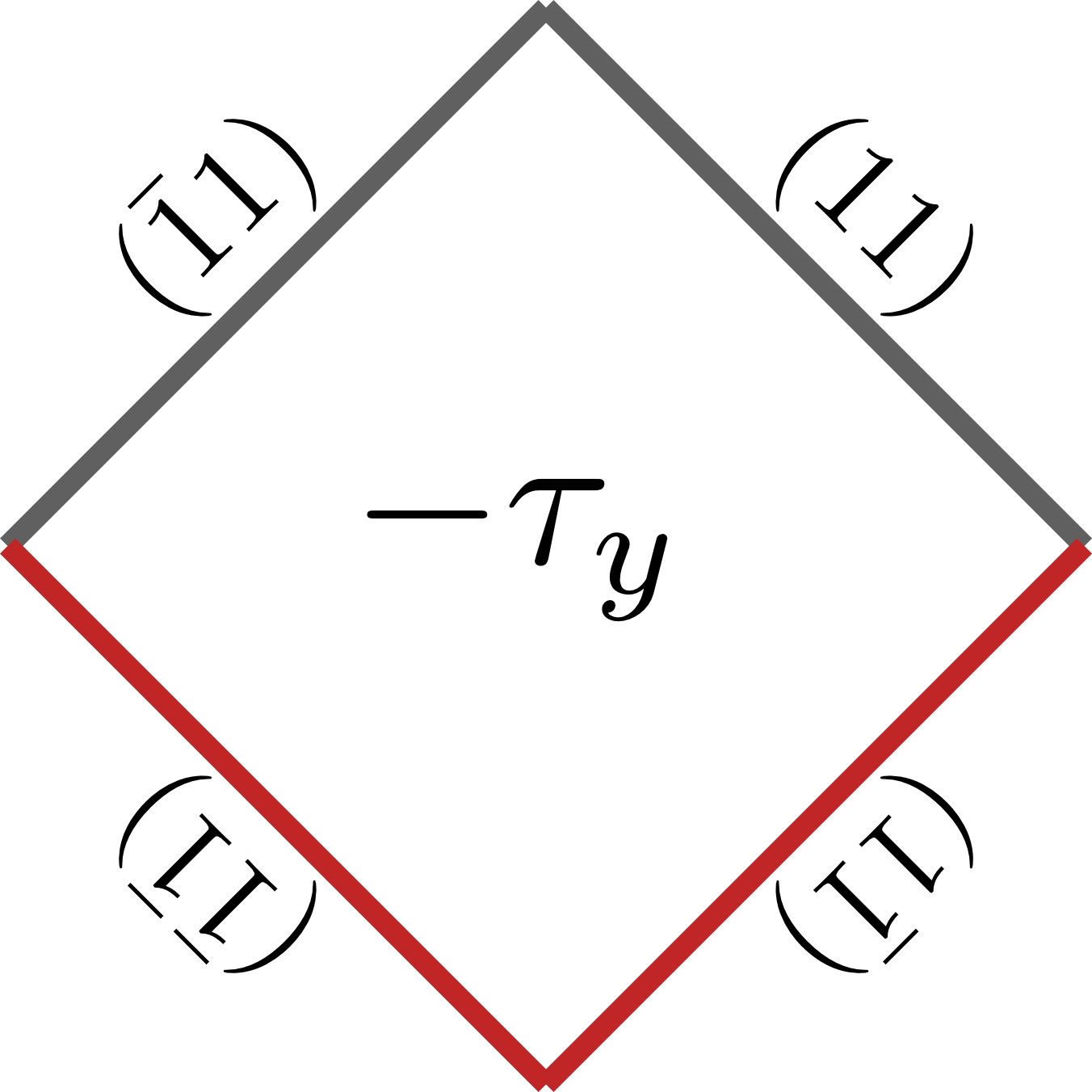}\label{fig:ic2z_surf_b1}}
\subfigure[]{\includegraphics[bb=0 0 402 455,width=4.0cm]{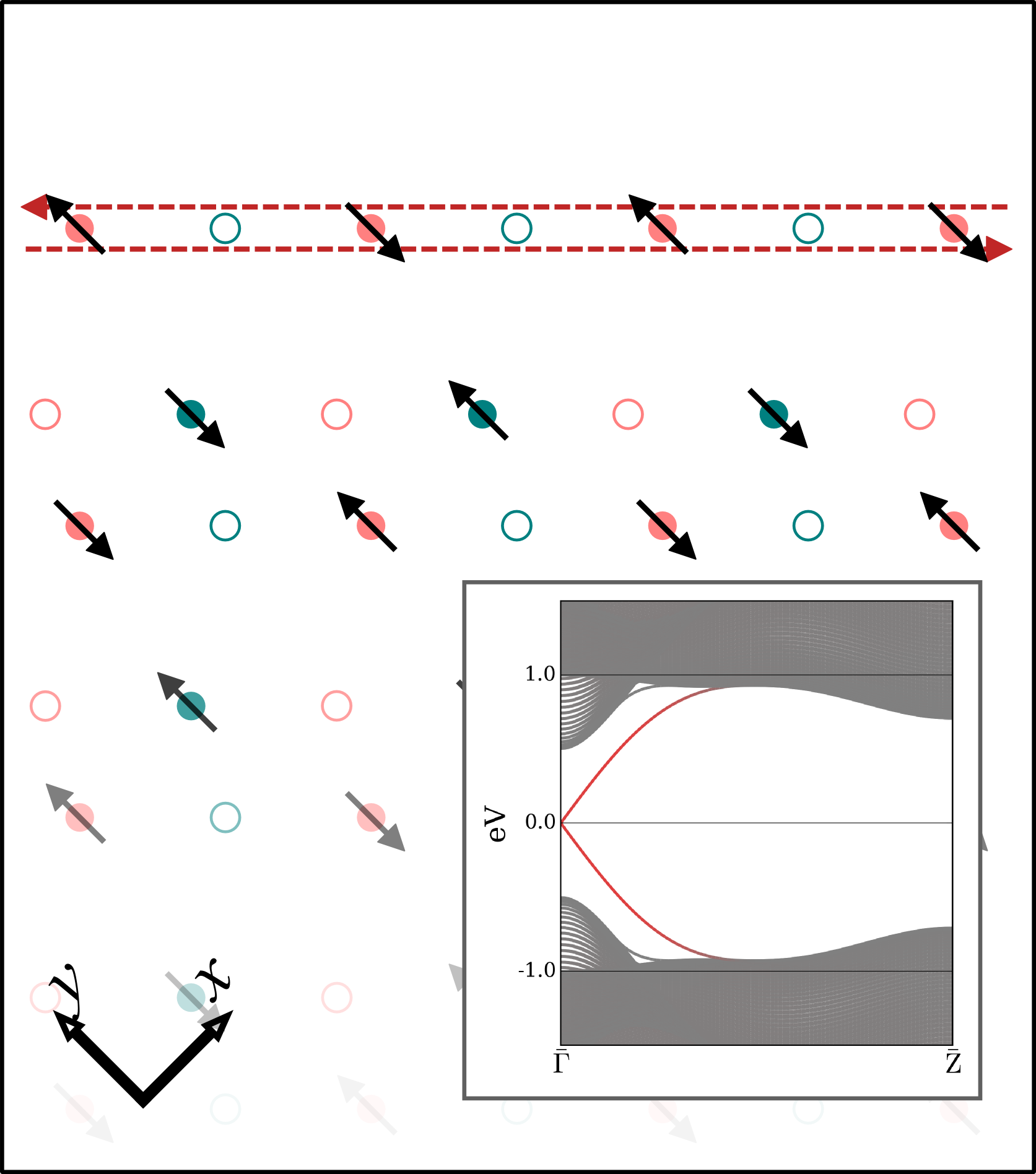}\label{fig:ic2z_surf_a2}}
\subfigure[]{\includegraphics[bb=0 0 402 455,width=4.0cm]{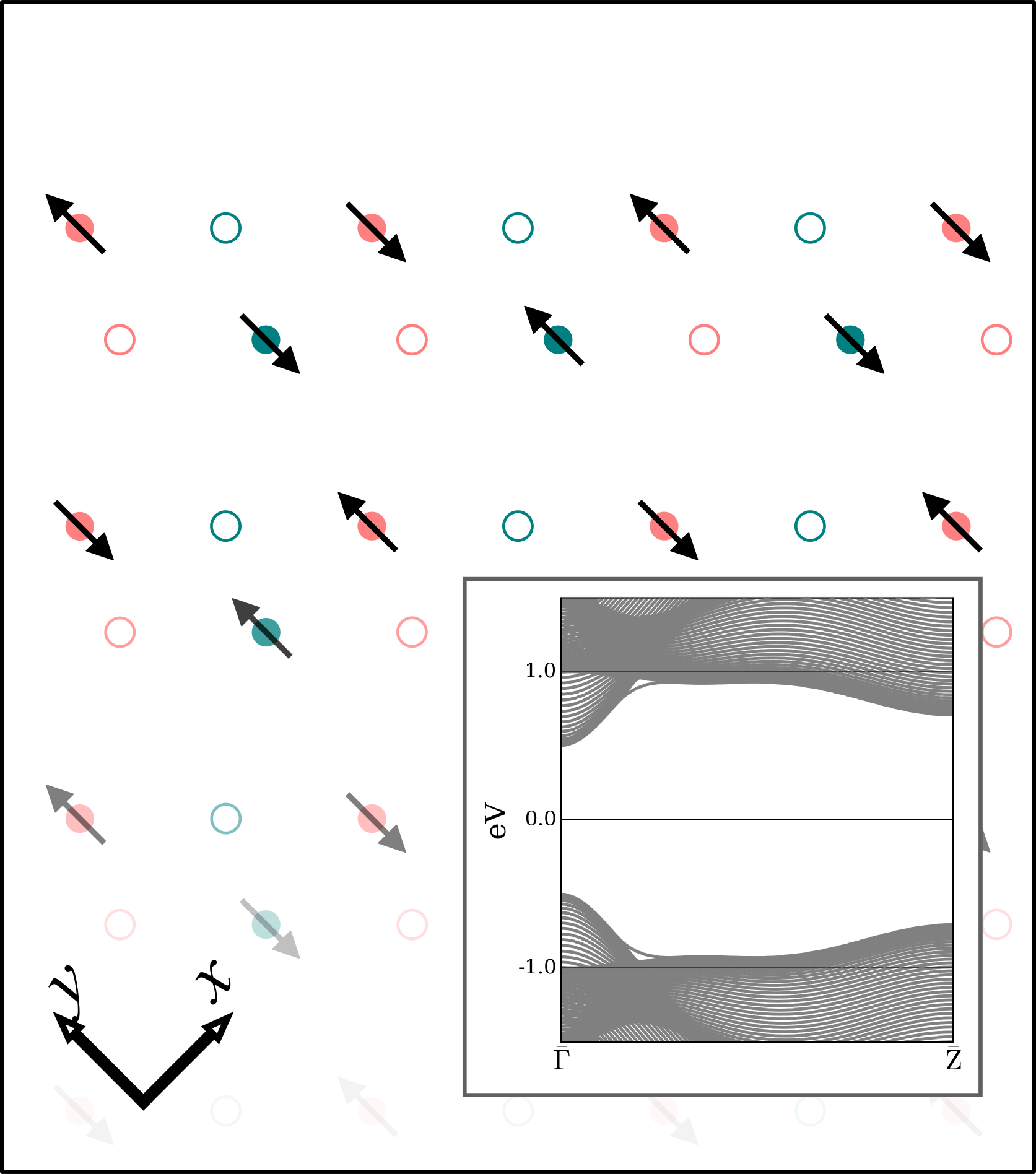}\label{fig:ic2z_surf_b2}}

\caption{For the tight-binding model in Eq.~\eqref{eq:tb1}, introducing an asymmetry in the hopping between $A,A'$ and $B,B'$ sites may result in edge states on $\bar{\mathcal{T}}$-preserving surfaces. \subref{fig:ic2z_surf_a2},\subref{fig:ic2z_surf_b2} The (11) edge of a ribbon with different signs of the $M_z$-breaking term $\tau_y$ in the $\steveBk\cdot\steveB{p}$ at $M$.  The edge in \subref{fig:ic2z_surf_a2}, represented by red lines in \subref{fig:ic2z_surf_a1} and \subref{fig:ic2z_surf_b1} hosts surface states, shown in the inset band structure. The edge in \subref{fig:ic2z_surf_b2}, represented by black lines in \subref{fig:ic2z_surf_a1} and \subref{fig:ic2z_surf_b1}, is fully gapped. Flipping the sign of $\tau_{y}$ is equivalent to applying $C_{2z}$ rotation, which exchanges the two crystalline phases.}  
\label{fig:ic2z_surf}
\end{figure}

To characterize this Dirac point as a quantum critical point, we examine symmetry-lowering perturbations that gap to insulating phases. In the $\steveBk\cdot\steveB{p}$ theory, five matrices preserve $\bar{\mathcal{T}}$ while breaking one of the spatial symmetries: $\tau_y$, $\tau_x\sigma_x$, $\tau_x\sigma_y$, $\tau_x\sigma_z$, and $\tau_{z}$.  Adding mass terms proportional to these matrices results in either insulating or Weyl semimetal phases, depending on the band ordering elsewhere in the BZ. Unlike in time-reversal-symmetric Dirac semimetals, the resulting gapped phases in these systems cannot be evaluated by a $\mathbb{Z}_{2}$ quantum-spin-Hall (QSH) invariant.  Furthermore, as $\bar{\mathcal{T}}^{2}=-1$ on only a line in the bulk BZ, this system also cannot realize the inherently 3D antiferromagnetic topological insulating phase described by Mong, Essin, and Moore in Ref.~\citenum{AFTI}. However, we find that this does not exclude the presence of 2D topological magnetic crystalline phases, i.e., those surface protected by $\bar{\mathcal{T}}$.

\begin{figure}
\centering
\subfigure[]{\includegraphics[bb=0 0 440 200,height=1.5cm]{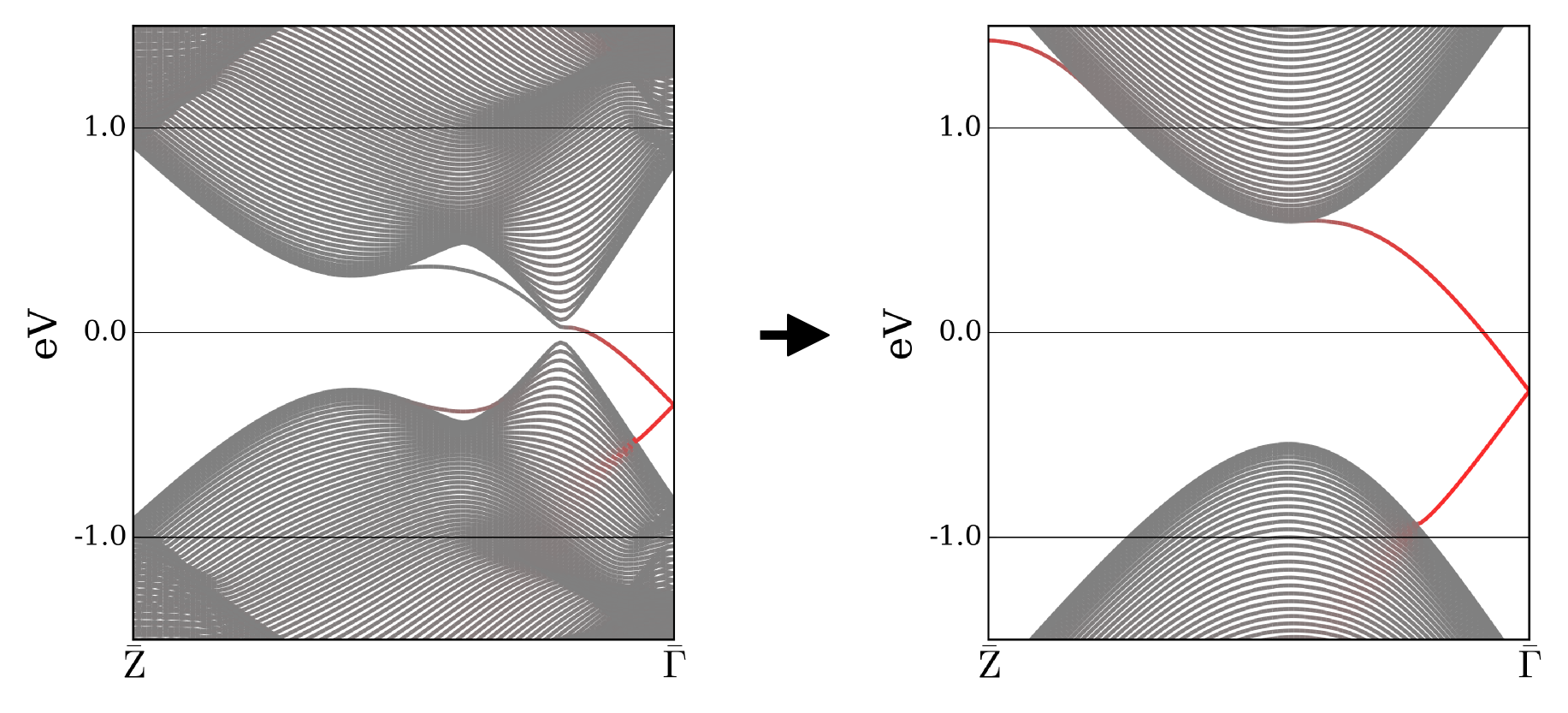}\label{fig:surf_ic2z_xx}}
\subfigure[]{\includegraphics[bb=0 0 664 200,height=1.5cm]{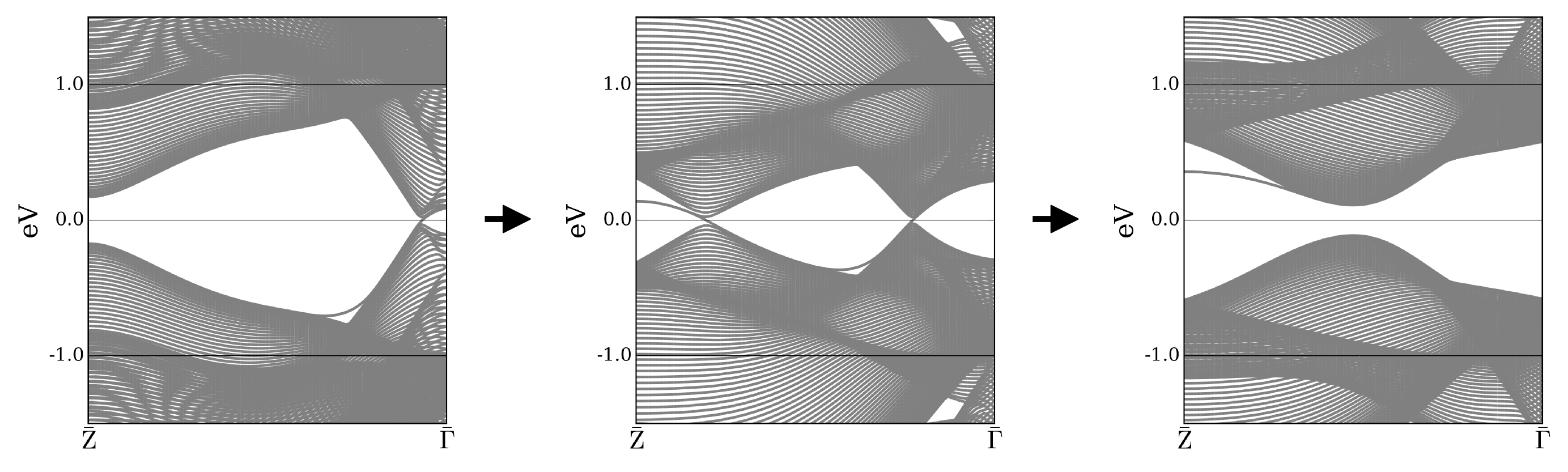}\label{fig:surf_ic2z_zz}}
\subfigure[]{\includegraphics[bb=0 0 720 360,width=4.25cm]{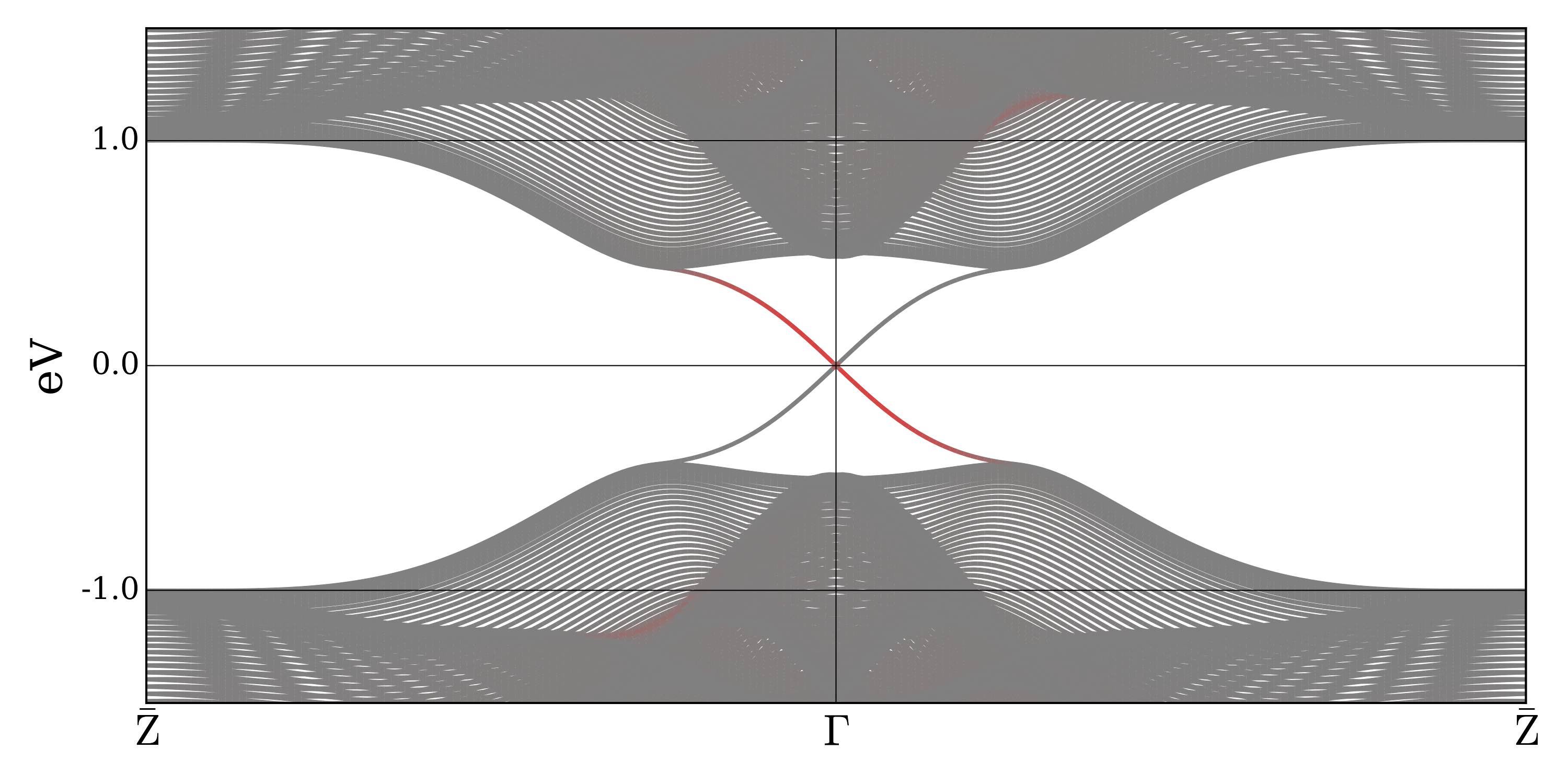}\label{fig:surf_ic2z_3}}
\subfigure[]{\includegraphics[bb=0 0 720 360,width=4.25cm]{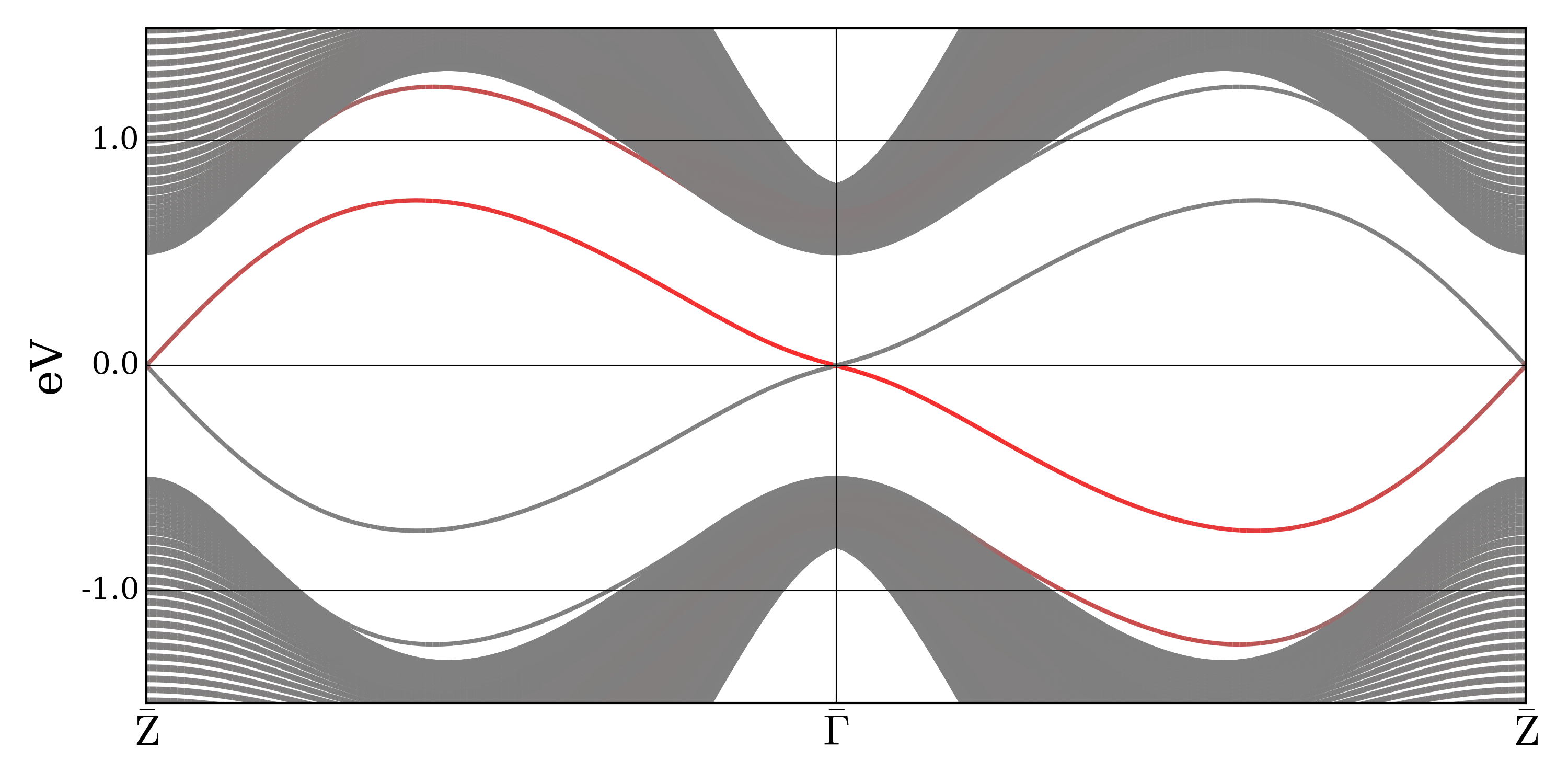}\label{fig:surf_ic2z_01}}

\caption{\subref{fig:surf_ic2z_xx} Perturbations corresponding to $\tau_x\sigma_x$ (along with $\tau_x\sigma_y$ and $\tau_x\sigma_z$) result in nodal phases (left) or bulk gapped phases (right), depending on perturbation strength. These cases are associated with dimerizations of the lattice, analogous to those in Fig.~\ref{fig:ic2z_surf}, and may produce edge states in the same fashion \subref{fig:surf_ic2z_zz}. $\tau_z$ represents a staggered on-site potential and leads to a pair of Weyl points for small magnitudes (left), two pairs as the magnitude increases (center), and, ultimately, an insulating phase once the Weyl points annihilate (right), but never produces edge states, independent of termination. \subref{fig:surf_ic2z_3} The chiral edge states resulting from breaking $\bar{\mathcal{T}}$ while preserving the spatial symmetries; left- and right-moving states sit on opposite edges and connect the valence and bulk manifolds, consistent with bulk topology $\abs{C}=1$.  \subref{fig:surf_ic2z_01}  Band structure of the (10) $\tilde{\mathcal{T}}$-breaking edge of the $\bar{\mathcal{T}}$-bulk-preserving perturbed system in Fig~\ref{fig:ic2z_surf}.  Each edge hosts a single, directional, trivial edge state, indicating that the $\bar{\mathcal{T}}$-preserving bulk-insulating phases are Chern-trivial $C=0$.}  
\label{fig:ic2z_surf2}   
\end{figure}

Consider the  $\left(11\right)$, $\left(1\bar{1}\right)$, $\left(\bar{1}1\right)$, and $\left(\bar{1}\bar{1}\right)$ edges, which preserve $\bar{\mathcal{T}}$. While one surface TRIM, $\bar{\Gamma}$, has a Kramers degeneracy from $\bar{\mathcal{T}}^{2}=-1$, the other TRIM, $\bar{Z}$, does not.  Inducing a distortion potential proportional to the mass term $\tau_y$ results in a bulk-insulating phase, and, for appropriately chosen terminations, edge states resembling QSH edge modes appear in the gap and meet in a linearly dispersive Kramers pair at $\bar{\Gamma}$ [Fig.~\ref{fig:surf_ic2z_3}].  The surface bands at $\bar{Z}$, however, are singly degenerate and free to move, as $\bar{\mathcal{T}}^2=+1$ and Kramers theorem is not enforced. This $\bar{\mathcal{T}}$-preserving gapped system behaves like an array of Su-Schrieffer-Heeger chains: $\tau_y$ effectively dimerizes the $A$ and $B$ sublattices, leaving an edge state on unpaired terminations.  Changing the sign of $\tau_y$ causes dimers to switch partners, converting edge states between paired and unpaired [Fig.~\ref{fig:ic2z_surf}].  The terms $\tau_x\sigma_x$, $\tau_x\sigma_y$, and $\tau_x\sigma_z$ also correspond to dimerizing distortions, and produce the same behavior as $\tau_y$, though when only weakly applied they result in Weyl nodes (or a nodal loop in the case of $\tau_x\sigma_y$, which preserves $M_z$) (Fig.~\ref{fig:surf_ic2z_xx}).   Finally, the term $\tau_z$, which corresponds to a staggered on-site potential, is only capable of gapping to a trivial insulator, though, when weakly induced, it also produces an intermediate Weyl semimetal phase [Fig.~\ref{fig:surf_ic2z_zz}].  

We can separate crystalline insulating effects from the overall bulk topology by examining the (10) $\tilde{\mathcal{T}}$-breaking edge in the bulk-$\tilde{\mathcal{T}}$-preserving insulating phases above.  Though inducing some of the previous mass terms leads this low-symmetry edge to display chiral modes [Fig.\ref{fig:surf_ic2z_01}], these modes do not connect the bulk and valence manifolds, and are, therefore, non-topological.  This indicates that the antiferromagnetic crystalline insulating phases in Fig.~\ref{fig:ic2z_surf} are Chern trivial $(C=0)$.  

The magnetic Dirac point can also be gapped by breaking $\bar{\mathcal{T}}$.  Applying a mass term $\sigma_{y}$ at $M$ breaks $\bar{\mathcal{T}}$ while preserving both spatial symmetries and results in the development of a single topological chiral mode on each edge [Fig~\ref{fig:ic2z_surf2}(c)], implying a Chern-insulating bulk with winding $|C|=1$.

\begin{figure}
\centering
\subfigure[]{\includegraphics[bb=-30 -30 220 220,height=3.5cm]{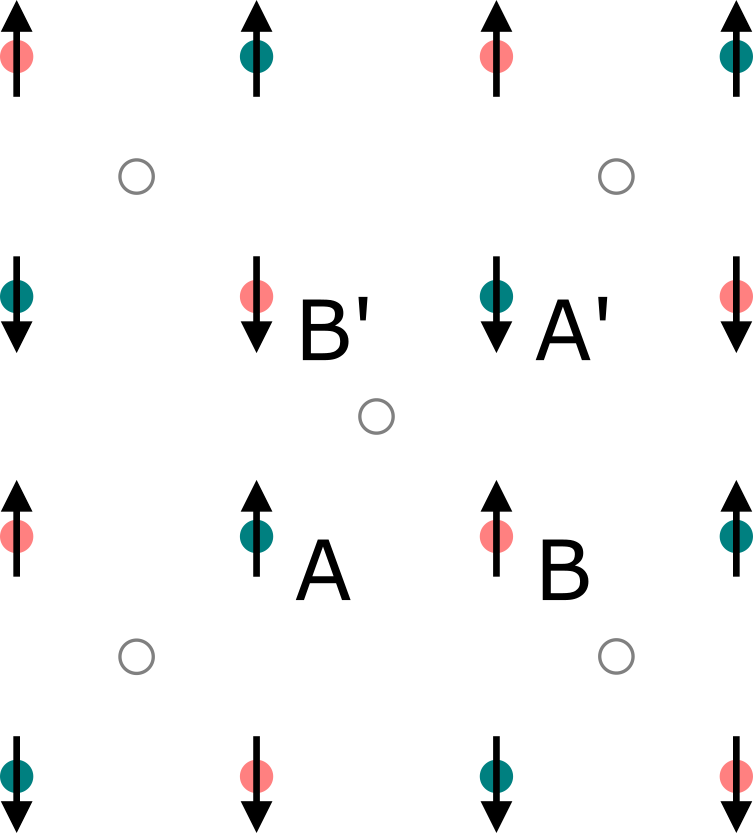}\label{fig:lattice_cc}}
\subfigure[]{\includegraphics[bb=50 0 600 400,height=3.5cm]{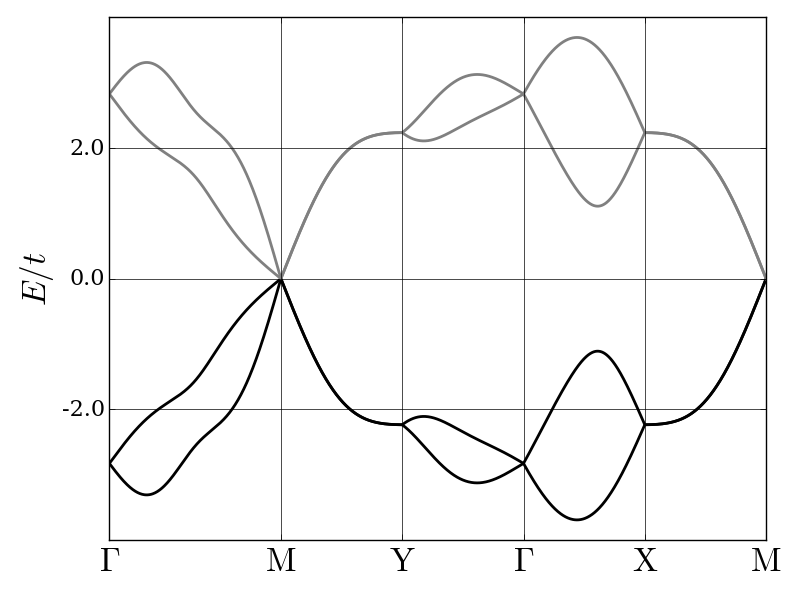}\label{fig:bands_cc}}

\caption{\subref{fig:lattice_cc} The lattice with $C_{2x}$, $C_{2y}$, and $\bar{\mathcal{T}}=\{\mathcal{T}| \frac{1}{2}  \frac{1}{2}\}$ with spin alignment $\pm\hat{y}$. The red and green sites sit above and below the plane, respectively; the gray open-circle site lies in the plane. The symmetries of this magnetic layer group require that bands, shown in \subref{fig:bands_cc}, while singly degenerate, still group together in multiples of 4 and meet at $M$ in Dirac points with nondegenerate cones.  Systems in this magnetic layer group are, therefore, filling-enforced magnetic Dirac semimetals at fillings $\nu\in4\mathbb{Z} + 2$.  These bands were obtained from a tight-binding model detailed in the Supplemental Material.}
\label{fig:cc}
\end{figure}

We also find that turning on the same antiferromagnetic potential in an otherwise symmorphic system also results in the enforcement of related magnetic Dirac points.  In Fig.~\ref{fig:lattice_cc}, we show a lattice generated only with antiferromagnetic time-reversal $\tilde{\mathcal{T}}$ and symmorphic rotations $C_{2x}$ and $C_{2y}$. At $M$ this combination of symmetries still satisfies the algebra in Eq.~\eqref{eq:alg}, and bands there consequently form similar magnetic Dirac points composed of two nondegenerate cones [Fig.~\ref{fig:bands_cc}].  Tight-binding models for this symmorphic system are detailed in the Supplemental Material.

For both magnetic Dirac systems, electron fillings $\nu\in4\mathbb{Z} + 2$ are required for the Fermi energy to lie at the Dirac point.  In the first model, the presence of multiple nonsymmorphic symmetries disallows fillings of $\nu=2,6$,  as they would imply atoms with fractional numbers of electrons.  However, derived phases for which one of the nonsymmorphic symmetries is broken are still achievable.  Appropriately chosen adatoms or substrates may be able to dope the system while only weakly perturbing it, maintaining an approximate Dirac cone.    

In the magnetic layer group of the symmorphic model in Fig.~\ref{fig:cc}, it is possible for pairs of sublattices to coincide, such that only two sites are necessary.  In such a two-site system, the Dirac point would be allowed to sit at the Fermi energy without the distribution of electrons violating crystal symmetries.  However, constructing a two-site model with these symmetries requires a more complicated pattern of magnetic ordering.

\begin{figure}
\subfigure[]{\includegraphics[bb=0 0 900 700,width=4.25cm]{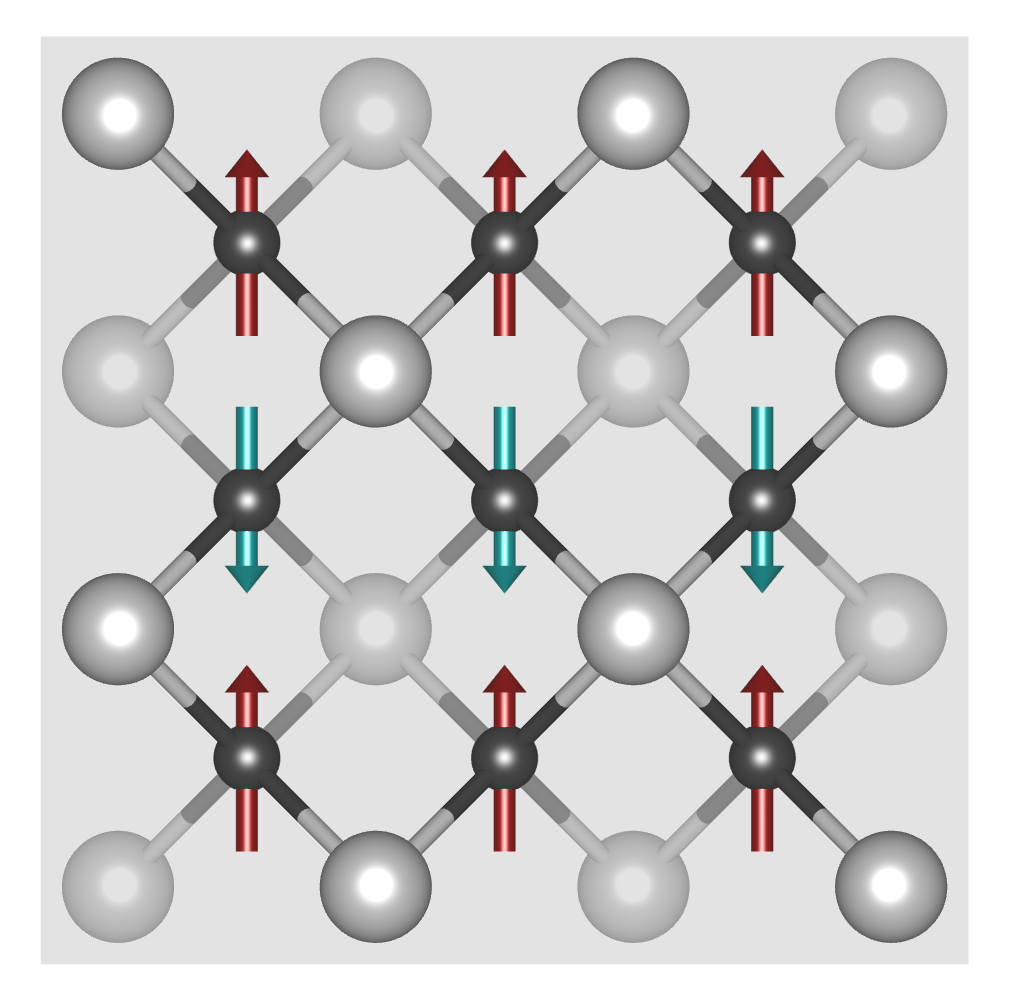}\label{fig:fese-struct}}
\subfigure[]{\includegraphics[bb=0 0 800 700,width=4.25cm]{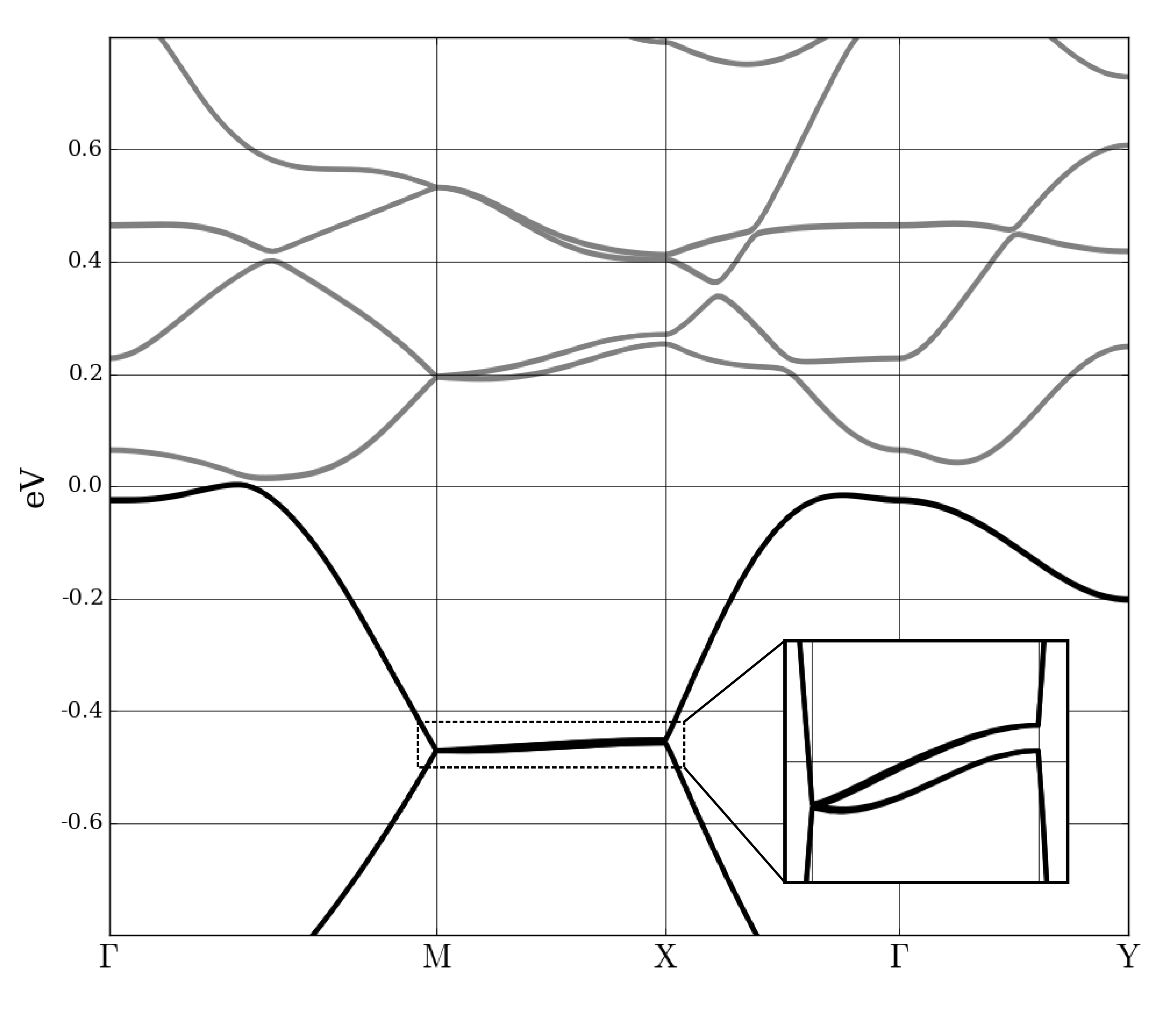}\label{fig:fese-bands}}

\caption{\subref{fig:fese-struct} The structure of an FeSe monolayer.  The iron atoms (dark gray) form a planar square lattice, while the selenium atoms sit above and below the plane, so that the iron atoms are tetrahedrally coordinated.  Magnetic moments are shown for the striped ordering phase, and are represented by the colored arrows. \subref{fig:fese-bands} The band structure of the striped phase of FeSe. Below the Fermi energy, the valence bands form a Dirac point at $M$ that splits weakly along the $M$-$X$ line.  The splitting is due to spin-orbit interaction and its weakness is a consequence of the bands comprising primarily iron $d$-orbitals.} 
 \label{fig:fese}
\end{figure}

This physics may be realized in the antiferromagnetic phase of iron-based superconductors.  The iron pnictides--and FeSe--comprise layers of iron arsenide or iron selenide in the antilitharge structure~\cite{Hosono2015}, and have already been shown to have nontrivial topological properties~\cite{Ran2009}.  Recently, monolayers of FeSe have been synthesized and investigated~\cite{Ge2014,Liu2012FeSe}. In the iron superconductors, including bulk FeSe, the antiferromagnetic order typically manifests as a striped pattern [Fig.~\ref{fig:fese-struct}]~\cite{Singh2008,Li2009,Yildirim2008,Glasbrenner2015,Wang2015} with symmetries captured in the $t_1^{\rm SO}=t_2^{\rm SO}$, $t_3^{\rm SO}=t_4^{\rm SO}$ limit of  Eq.~\eqref{eq:tb1}. Using a DFT calculation~\footnote{See Supplemental Material}, we obtain the band structure of FeSe (Fig.~\ref{fig:fese-bands}), which exhibits clear fourfold-degenerate Dirac fermions at $M$. For single-layer FeSe, the filling prevents the Fermi energy from sitting at any of the Dirac points.  However, by stacking iron pnictide monolayers with intercalated species, it may be possible to engineer a few-layer system with the correct filling.

We have described a class of magnetic Dirac semimetals protected by modifying time-reversal symmetry to include a fractional translation.  This translation results in commutation relations with the spatial symmetries different from those in ordinary time-reversal-symmetric crystals, allowing for Fermi surfaces consisting of single Dirac points, and removing the requirement of nonsymmorphic symmetries.  Both topologically nontrivial magnetic crystalline insulating and Chern insulating phases are easily accessible from this magnetic semimetallic phase by breaking symmetries.  The dimerizations required to gap into the $\bar{\mathcal{T}}$-preserving phases can, in general, be achieved by applying 11-direction strain, and provide a route towards strain-engineering broken-time-reversal quantum phase transitions.  We find that FeSe monolayers with striped magnetic ordering display these magnetic Dirac fermions.

The role of disorder and interactions in magnetic Dirac semimetals is an open question.  A given disorder ensemble may preserve magnetic group symmetries on the average while, nevertheless, representing a quantum critical point nonperturbatively related to the single-particle magnetic Dirac point, or the system may Anderson localize.  However, if the mean-field theory still obeys the group symmetries, Eq.~eq\ref{eq:alg} remains satisfied and a gap cannot form.  In strongly-correlated time-reversal-symmetric filling-enforced semimetals SrIrO$_3$ and CuBi$_2$O$_4$, there have been hints of Mott instability related to emergent spin order~\cite{IridateMott,NewFermions,DDPMott}.  As magnetic Dirac semimetals are already stable under spin ordering, their gaplessness may, therefore, be more robust against interactions.

Finally, we note that these magnetic semimetals circumvent the Dirac fermion doubling theorem for time-reversal-symmetric Dirac semimetals.  In those systems, unpaired Dirac points are prevented from being stabilized in 2D bulk crystals by the presence of additional Dirac or Weyl features at the Fermi energy.  This prevents the nearby QSH and trivial insulating phases from being related to each other by a crystal symmetry operation~\cite{Steve2D}.  Though the $\bar{\mathcal{T}}$-preserving gapped phases in our systems seem to violate this doubling theorem, they are actually unrelated.  The gapped phases in these magnetic  Dirac systems are topological crystalline phases preserved by time-reversal and a surface-specific spatial operation, here, the combined operation of time-reversal and a diagonal half-lattice translation.  The two antiferromagnetic topological crystalline insulating phases in Fig.~\ref{fig:ic2z_surf} are Chern-trivial and related by a $C_{2z}$ operation, such that only crystalline invariants are exchanged under spatial operations and the overall bulk topology remains unaffected.  In fact, one may, instead, consider the magnetic Dirac points presented here as the symmetry-pinned combinations of two, twofold-degenerate quantum Hall transitions.  In this sense, these Dirac points also successfully avoid the two-dimensional parity anomaly for twofold-degenerate fermions addressed by Haldane in Ref.~\citenum{HaldaneModel}.

\begin{acknowledgments}
We thank Charles L. Kane and Eugene Mele for helpful discussions. SMY was supported by a National Research Council Research Associateship Award at the US Naval Research Laboratory.  BJW acknowledges support through a Simons Investigator grant from the Simons Foundation to Charles L. Kane and through Nordita under ERC DM 321031.  During the final stages of preparing this manuscript, a band-inversion magnetic Dirac semimetal was presented in Ref.~\onlinecite{AFDIRAC}.
\end{acknowledgments}

\end{document}


\title{\bf Supplemental Material for Filling-enforced Magnetic Dirac Semimetals in Two Dimensions}
\author{Steve M. Young}
\affiliation{Center for Computational Materials Science, Naval Research Laboratory, Washington, D.C. 20375, USA}
\author{Benjamin J. Wieder}
\affiliation{Nordita, Center for Quantum Materials, KTH Royal Institute of Technology and Stockholm University, Roslagstullsbacken 23, SE-106 91 Stockholm, Sweden}

\maketitle

\section{Model for a Magnetic Dirac Point in a Fully Symmorphic Crystal}

\begin{figure}

\subfigure[]{\includegraphics[bb=0 0 400 360,height=1.55cm]{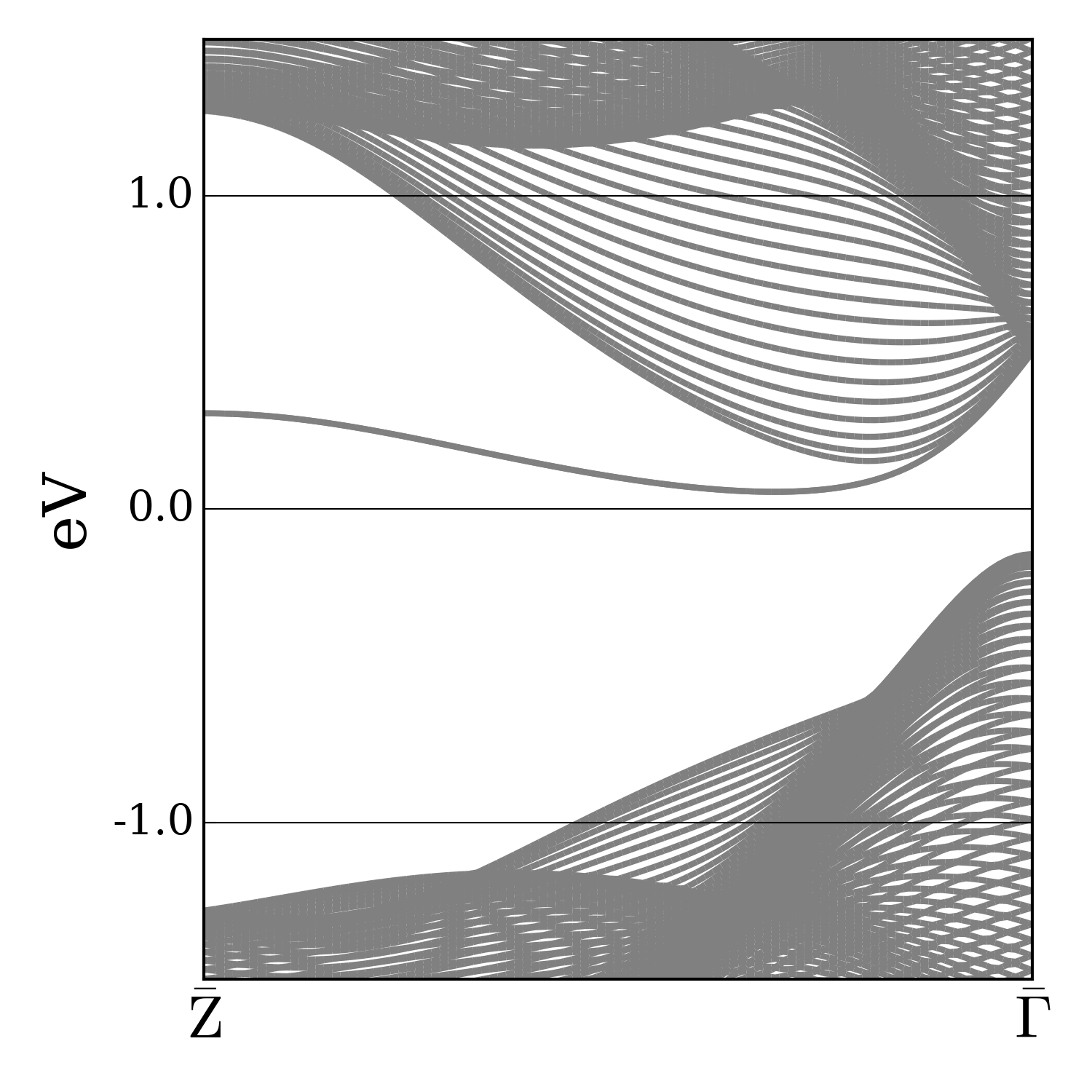}\label{fig:surf_cc2_1}}
\subfigure[]{\includegraphics[bb=0 0 440 200,height=1.55cm]{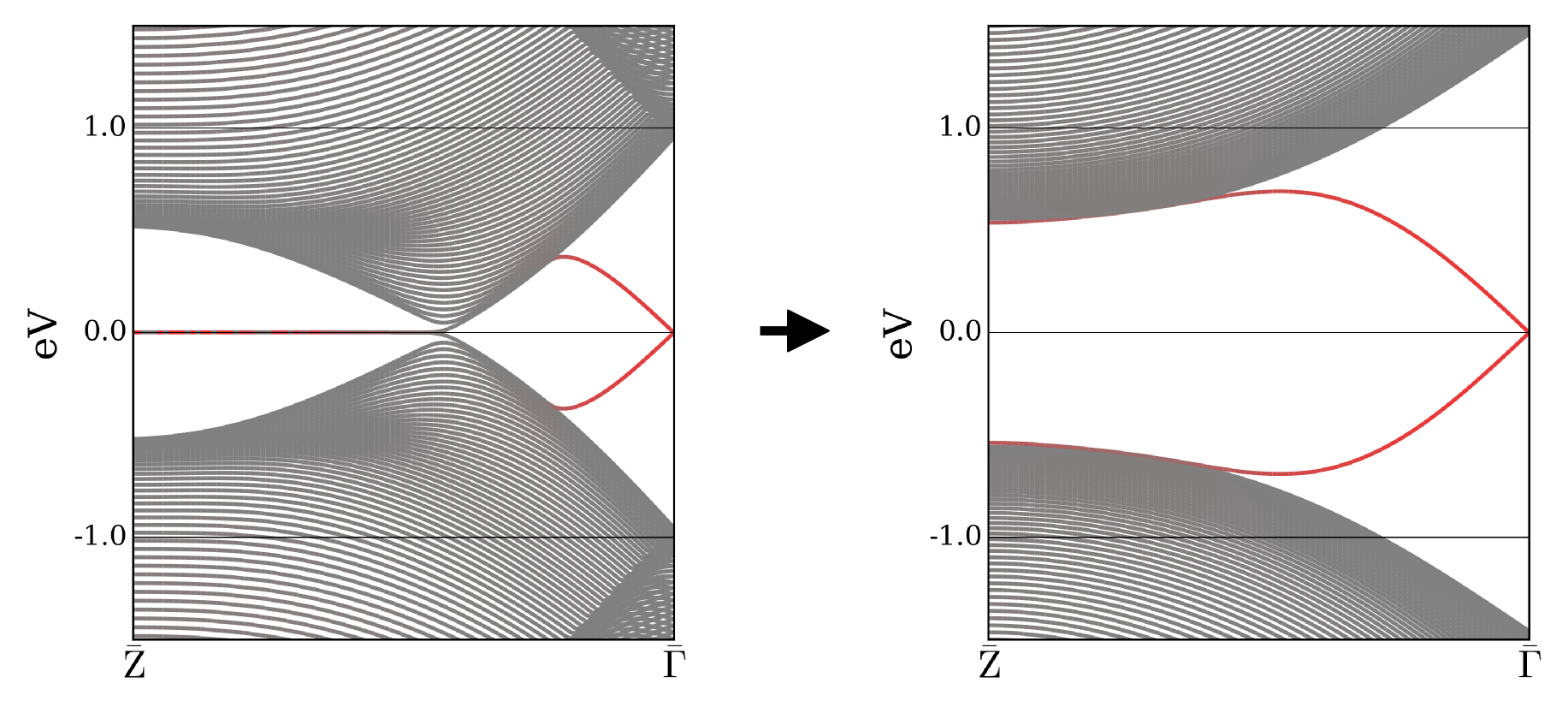}\label{fig:surf_cc2_2}}
\subfigure[]{\includegraphics[bb=0 0 720 360,height=1.55cm]{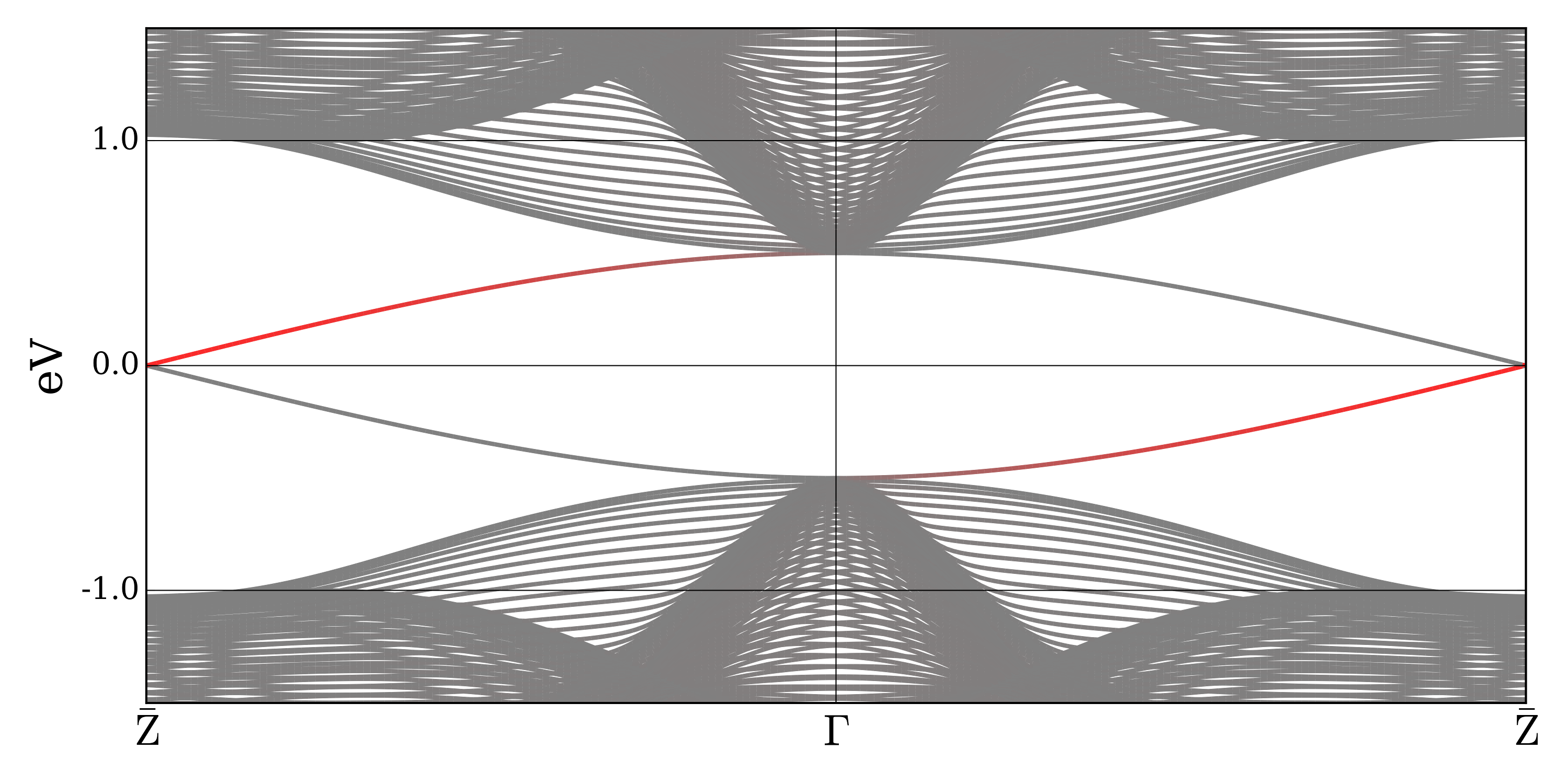}\label{fig:surf_cc2_3}}

\caption{The band structure of the $\langle 11\rangle$ edges of a ribbon of the system in Eq.~\eqref{eq:tb2} for~\subref{fig:surf_cc2_1} a staggered on-site potential and dimerizing interactions with~\subref{fig:surf_cc2_2} low and high magnitudes.  In the latter case, only the termination with unpaired sites (as in Fig.~2(c) of the main text) is represented. As before, paired edges yield no edge states. Band structures when the bulk is gapped by~\subref{fig:surf_cc2_3} a $\bar{\Theta}$-breaking term display topological chiral edge states associated with a Chern number $|C|=1$. }
\label{fig:cc2_surf}
\end{figure}

As noted in the main text, we find that unlike in time-reversal-symmetric crystals, filling-enforced Dirac points can be formed in magnetic crystals utilizing only symmorphic crystal symmetries.  Shown in Fig. 4(a) of the main text, we can construct a lattice using $C_{2x}$, $C_{2y}$, and $\tilde{\mathcal{T}}=\{\mathcal{T}|\frac{1}{2}\frac{1}{2}\}$.  At the corner TRIM $M$ $(k_{x}=k_{y}=\pi)$, its $\steveBk\cdot\steveB{p}$ theory looks like
\begin{flalign*}
&C_{2y}=i\tau_y,C_{2x}=i\tau_x\sigma_y,\bar{\mathcal{T}}=i\tau_z\sigma_yK\\
H_{\rm M}=&\left[t_0\tau_x+t_1^{\rm SO}\tau_z\sigma_z+t_2^{\rm SO}\tau_z\sigma_x\right] k_x\\
&+\left[t_1^{\rm SO}\sigma_z+ t_2^{\rm SO}\sigma_x +t_3^{\rm SO}\tau_y\sigma_y\right] k_y.
\end{flalign*}

We can also generate a full-BZ tight-binding model for this magnetic layer group:

\begin{flalign}
\mathcal{H}
=&t_0\cos\left(\frac{k_x}{2}\right)\tau_x+t^{\rm SO}_3\cos\left(\frac{k_y}{2}\right)\tau_y\sigma_y\nonumber\\
&+t^{\rm SO}_1\left[\sin\left(k_x\right)\cos\left(k_y\right)\tau_z\sigma_z+\sin\left(k_y\right)\cos\left(k_x\right)\sigma_z\right]\nonumber\\
&+t^{\rm SO}_2\left[\sin\left(\frac{k_y}{2}\right)\cos\left(\frac{k_x}{2}\right)\tau_z\sigma_x\right.\nonumber\\
&\hspace{2cm}\left.+\sin\left(\frac{k_x}{2}\right)\cos\left(\frac{k_y}{2}\right)\sigma_x\right]
\label{eq:tb2}
\end{flalign}
for which we have enumerated \emph{all} symmetry-allowed terms up to second-nearest-neighbor hopping.  One possible band structure is shown in Fig. 4(b) of the main text.  

As in the first model in the main text, $\tau_z$ corresponds to an on-site potential that breaks the symmetry between the A and B sublattices, and $\tau_y$, $\tau_x\sigma_x$, $\tau_x\sigma_y$, and $\tau_x\sigma_z$ correspond to potentials that dimerize the A and B sublattices and produce crystalline edge states when pushed through an intermediate Weyl semimetal phase (Fig.~\ref{fig:surf_cc2_2}). Inducing $\tau_z$ instead opens a gap without an intermediate Weyl semimetal phase (Fig.~\ref{fig:surf_cc2_1}). In this symmorphic system, breaking $\bar{\mathcal{T}}$ can also result in $|C|=1$ chiral edge states (Fig.~\ref{fig:surf_cc2_3}).

\section{Density Functional Theoretic Methods}

In the main text, we present band structures for FeSe in a phase with striped magnetic ordering.  We obtain band structures for this material,shown in Fig.~5(b) of the main text, through the use of Density Functional Theory (DFT).  We performed DFT calculations using ~{\small QUANTUM ESPRESSO} with norm-conserving pseudopotentials generated using OPIUM~\cite{Rappe90p1227, Ramer99p12471} at the level of PBE. An energy cutoff of 50Ry was used, with a $24\times 24\times 1$ k-point grid and 15\AA~of vacuum.  

%